\documentclass{article}
\begin{document}

 \qquad \qquad STATISTICAL PHYSICS AND DYNAMICAL SYSTEMS:

\qquad\qquad \qquad  MODELS OF PHASE TRANSITIONS

\qquad   \qquad  \qquad \qquad \qquad Ajay Patwardhan

  Physics department, St Xavier's college, Mahapalika Marg, Mumbai, India

\qquad \qquad  Visitor, Institute of Mathematical Sciences, Chennai, India

\qquad \qquad \qquad \qquad \qquad \qquad  ABSTRACT

This paper explores the connection between dynamical system properties and statistical phyics of ensembles of such systems. Simple models are used to give novel phase transitions; particularly for finite N particle systems with many physically interesting examples.

\qquad \qquad \qquad \qquad  1. INTRODUCTION

The developments in dynamical systems and in Statistical Mechanics have occured for over a century. The ergodic hypothesis of J W Gibbs and Boltzmann's H theorem were re- investigated as integrable and chaotic dynamical systems were found. Poincare, Birkhoff, Krylov formulated these problems. Kolmogorov, Arnold and Moser theorem gave a detailed description of phase space as 'islands of integrable and sea of chaotic regions'.

 From chaotic billiards to integrable Toda lattices the range of the systems includes mixed type of systems. Work of Ruelle, Benettin, Galgani, Casati, Galavotti and others clarified some properties of classical and quantum systems. Recent work of E.G.D.Cohen, L Casetti and others has given a rigourous basis for Riemannian space based description of dynamics of Hamiltonian systems and K entropy. However the Statistical mechanics of dynamical systems remains to be formulated.

 Ensembles with model Hamiltonians showing chaos transitions and topological transitions have become significant as simulation and empirical work has grown. A partition function inclusive of the Kolmogorov entropy and the Euler characteristic has been defined. Nano clusters of particles show properties dependent on the boundaries, number, energy and interaction parameters. Strongly interacting systems in condensed matter , nucleons and quarks are also possible applications.

 This requires that any dynamical system model such as with maps, or differential equations for the 'units' in the physical system, will have consequences for the statistical mechanics of their ensemble. It leads to novel phase transitions and new interpretations of thermodynamic phase transitions. This has been shown in this paper for some simple models .

 The Baker transform is used to model kinetics of  melting in two dimensions. The Henon Heiles like models are used to model ergodic channels with correlated charge densities in superconductors. The gas of molecules with Henon Heiles Hamiltonian is shown to have a 'phase' transition dependent on the chaotic transition in the molecule. Quantum chaos also creates a transition in the ensemble of such systems and the Poisson, GOE, GUE and Husimi distributions are an example of Wigner distributions on phase space.

 These ideas could be generalised to more complex model maps or Hamiltonians, that are used in physical systems. Hence a statistical physics of finite ( any N ) number of particles is expected to have definite properties dependent on the integrable to chaotic transitions in their units. This can be seen in the Toda and Fermi Pasta Ulam like systems. The description of Hamiltonian systems as flows on Riemann spaces gives a intrinsic definition of geodesic deviation equation and Lyapunov exponents. This has led to defining the connection between statistical mechanics and dynamical systems in a fundamental way.

2.  MODEL FOR KINETICS OF MELTING AND FREEZING IN TWO DIMENSIONS , USING BAKER LIKE TRANSFORMS

The folding property of this transform causes mixing and ergodicity and has a K entropy. Any regular structure of points in the $(0,1)$ square, will after many iterations become 'smeared' all overthe square. A two dimensional crystal , a snow flake, a liquid crystal, a spin or metallic glass has a kinetics of melting and freezing. A order parameter and correlations with a time scale dependent on rate of cooling and heating are present. Any model for the kinetics of melting , converted into a difference equation with a time step and  a folding with two subintervals in a square is like a Baker transform.

Consider this process modeled by a Baker transform with a time step for iteration and a 'unit cell square'. 

A point $  x_{n}$, $ y_{n}$ goes to $ x_{n+1}$ , $ y_{n+1} $ 

using matrix mapping $  (1,1)$,$(1,2) $ with this 'cat map' having eigenvalues

 $ 0.5(3+/- \sqrt5)$ = $ exp(+/-\sigma)$. 

This gives the Lyapunov exponent $\sigma$. Starting from any initial point in the square $ n $ iterates will give the randomised distribution over a ordering length scale

 $l_{max} = \tau^{-1/\sigma} $ ; where $\tau$ is the time step and the $\sigma$ is the K entropy.

A more general model would use different maps on the half intervals. 

Diagonal $(2,0.5)$ matrix for lower half $0<x_{n}<0.5$ ; 

and the same matrix acting on $x_{n}$,  $y_{n}$ with $(-1,0.5)$ deducted on the interval $0.5< x_{n}<1$. 

More generally the Baker like transforms can be taken as

 $x_{n} <\alpha <1 $ and for $x_{n}> \alpha < 1$. 

Then $ y_{n+1}= \lambda_{a} y_{n}$, $ x_{n+1} =x_{n}/ \alpha$ and respectively,

 $y_{n+1} = \lambda_{b} y_{n} +0.5$ ,$ x_{n+1} = \frac{x_{n} - \alpha}{1-\alpha}$

 for $\alpha$, $\lambda_{a}$, $ \lambda_{b} $ all between $ 0 $ and $1$. 

This gives a number of adjustable parameters to model a variety of melting and freezing in two dimensions.

Two point correlations can be found

 $< \phi(\omega_{1}) \phi(\omega_{2}> =  <\phi(\omega_{1})><\phi(\omega_{2})> $

 for two regions $\omega_{1} $ and $ \omega_{2} $ in the square. These can give parametrisation in terms of experimentally observed values.

 From exact crystalline symmetry to random network of bonds ,  the iterated map over the time interval of melting or freezing gives a model for partial mixing. Consider each unit cell modelled by the Baker like transform; and a range of parameters that can vary across the sample. Coordination clusters, ionic mobility, cooling or heating rates and correlation lengths are all measured quantities which are related to the parameters of the Baker like transforms.

 The dynamical phase transitions can create a variety of configurations. Equilibrium partition functions are defined at beginning and end stage . But rapid cooling or heating leads to multiple energy minima and entropy maxima , with the statistical entropy ( Kolmogorov entropy ) playing a role in the thermodynamic entropy. Ordered and disorderd states are formed at intermediate times, with transitions among them. The basic quantity, Lyapunov exponent of the map or dynamical system is connected to the basic quantity of the condensed matter system , the correlation length. The ensemble of ' cells ' with the Baker mapping iterated on each is a model of the  kinetics of melting and freezing in two dimensions.

3. MODEL FOR ERGODIC CHANNELS WITH CORRELATIONS, IN HENON HEILES LIKE SYSTEMS, AND SUPERCONDUCTIVITY .

The phase space has  islands of integrability and chaotic sea picture. Ergodic channels can form on repeated or periodic lattice structures, that create connected regions of the chaotic sea. In these regions two point correlations can be non zero. Consider a 'unit cell' with ergodic regions that are connected to those in neighbouring 'unit cells'. 

Then over some order parameter scale there is a continuous connected chaotic region. In this ergodic channel , across the sample there are non zero correlations. This could represent a model of the axial and planar degrees in a unit cell of a high temperature superconductor modeled by Henon Heiles type of Hamiltonian for the electron.

The Henon Heiles (HH) like Hamiltonian :

 $ H = 0.5(p_{1}^{2} + p_{2}^{2}) + 0.5 (q_{1}^{2} +q_{2}^{2}) +\alpha q_{1}^{2} q_{2} - \frac{\beta}{3}q_{2}^{3} $

 with $\alpha =1$ and $\beta= 1$ for the original HH case. 

For $ H=E$ with $E<E_{c1} =\frac{1}{12}$ phase space is mostly periodic or quasi periodoc motion.

 For $ E > E_{c2} = \frac{1}{6} $ it is mostly chaotic 

and for in between energies it has  mixed chaotic and integrable subspaces. The single connected ergodic region has a Lyapunov exponent $\sigma = 0.03$.

 Consider the axial and planar direction coordinates for the electron to be $q_{1}$ and $q_{2} $ for the unit cell in a orthorhombic structure as occuring in high $ T_{c}$ superconductors. While the phase space is mostly integrable the electron is bound in cell; however if the mixed form occurs the electron can traverse the chaotic sea component, which increases in its volume fraction as chaotic transition occurs. The contiguous ergodic channels in neighbouring unit cells connect to form a sample wide ergodic region.

 A correlation length scale or order parameter can be found. For an intermediate energy $ E = 0.125 $ the phase space has half volume integrable and half chaotic. The area fraction computed numerically as a function of energy is a straight line. The broken or partial ergodicity on phase space requires a weighted average over a disjoint union of regions with different ergodic properties, such as the Kolmogorov entropy. Oseledec and Pesin's work gives   entropy definition as sum of K entropies, and an invariant measure for integration.

Conductivity arises in this model by electron motion in chaotic sea channels in classical case and in connected Husimi probability distribution in quantum case. 

$< \rho_{\omega_{1}}  \rho_{\omega_{2}} > $ is non zero

$< \frac{d}{dt} \rho_{\omega_{1}} \frac{d}{dt}  \rho_{\omega_{2}} > $ is non zero

for the probability density $\rho $ in the two regions $\omega_{1} $ and $ \omega_{2} $

$\frac{d}{dt}\rho = [\rho , H] $ brings the explicit model dependence from the Hamiltonian.

Hence the charge and current densities are correlated  at two points in the sample. If this ergodic channel extends over the whole sample and is continuous , then the correlation length scale allows an effective resistance less transfer of charge and current fluctuations from any point in sample to any other ; hence superconductivity occurs. The property is seen as a consequence of the nonlinearity rather than the Cooper pairing mechanism. 

The condition for this transition from conductivity to superconductivity to occur can be obtained in terms of the Jacobian $ J (\omega_{1} , \omega_{2} )$ by expanding $\rho $ to first order in $\omega$. This should leave the non zero correlation condition unchanged , that is the variation in the two point correlation is zero.

 Then the correlation length order parameter $\zeta $ can be defined in terms of the Jacobian. A Fokker Planck like equation in the ergodic channel for a two particle distribution for  $\rho $ can be defined. The small spread in $ T_{c} $ in the resistance versus temperature data in high $ T_{c}$ superconductors  could be attributed to the variation and number of ergodic channels available in parallel in the sample.

 The mixed state, 'dirty' or granular superconductors depend on the details of the microstructure to obtain coherence length, whereas long range correlations are introduced by ergodicity in this approach. However the $ T_{c} $, critical magnetic fields, energy gap and currents are not easily obtained in terms of the nonlinearity or chaotic transition energy surfaces in this model.

Charge and current correlations rather than transport, in specialised regions in classical or quantum phase space, and its projection onto real space for the mechanism of conductivity and superconductivity need further work. Model hamiltonians and energy and parameter ranges, creaing ergodicity occur frequently in dynamical systems;  which will have consequences for the statistical physics of condensed matter.

4. QUANTUM CHAOS AND STATISTICAL PHYSICS

The phase space version of quantum mechanics gives Wigner distributions. Many computed systems are known with chaotic transition to Husimi distributions; with connected and disconnected regions and probability measures on them. This implies that for models which have ensembles of such dynamical systems the statistical physics must have averaging on the distributions, which show a chaotic transition dependent on parameters . Hence a phase transition for the whole system depends critically on this chaotic transition in its constituents.

 In the case of energy level spectra the distribution of these energy levels is given by a GOE or GUE distribution for chaotic case and a Poisson or Wigner like one for the integrable case.

The energy distribution probability is given generally  by:

$P(E)= a[ \frac{E}{<E>}]^{\alpha} Exp( -[\frac{E}{<E>}]^{\beta}] )$

 is in chaotic case and it goes to $ \alpha = 0 $ and $\beta = 1 $ in integrable Poisson or Wigner case. $ <E> $ is the average energy and $a $is positive, and $ \alpha$ and $\beta$ are positive exponents, typically equal to two in the chaotic GOE, GUE cases. The phase transition depends on the parameter that switches the energy distribution from chaotic to integrable case.

 Any partition function statistical physics average should include an additional multiplicative weight or measure; that assigns probabilities for accessing the energy levels, as given by these distributions. And the parametric transition to chaotic distribution will  reflect in the partition function and its partial derivatives that give thermodynamic quantities. This is a phase transitionof a new kind or an alternate interpretation of usual phase transitions ; that is debatable.

This is an intrinsic effect, not dependent on the thermal reservoir, but dependent on the hamiltonian and its parameters. In an ensemble of systems with variation in hamiltonian and its parameters allowed this effect will be seen. Often the precise hamiltonian and its parameters and form is not known for a sample and hence it is essential to take a ensemble of these and average over them. So does this entail a modified description of canonical and grand canonical ensembles or a new ensemble; that remains an open question. If a bulk system is a collection of nano systems then such an ensemble may be defined to relate nano to bulk sample properties.

Possibly the best examples for these distributions arise in quantum optics where radiation - matter interactions occur. A non thermal like  radiation spectrum will result if the $ P(E) $ function is used in averaging. It will also show a parametric transition. Anharmonic oscillators in equilibrium with radiation can be experimentally observed to show this behaviour. Condensed matter examples in which the density of states function is modified by this $ P (E) $ multiplying the usual onecan show a parametric rather than thermal effect, in the chaotic transition. In extended and localised states, the density of states function will have this $P(E)$ function as  a multiplier.

5. TODA AND FERMI PASTA ULAM LIKE MODELS

Toda models have near neighbour interaction with a  exponential potential. They are known to be integrable. While Fermi Pasta Ulam models have polynomial potentials ( typically quartic) and show a variety of phenomena dependent on energy, number and parameters. The dynamics of a chain of coupled masses with a general potential $ V(x)= \sum a_{n} x^{n} $ can in the limit, show a Boussinesq like equation, and using quadratures method give a solitary wave solution. 

A typical potential will be polynomial plus exponential if it is of  Fermi Pasta Ulam plus Toda type. Simulations of the dynamics of such a chain can be shown to have a chaotic transition for reasonably small number of particles $ (< 20)$ and energies. For a range of parameters such as interparticle separation, and coupling parameters occuring as coefficients in the potential, this is true.

 This system was a prototype for the question : how does classical mechanics become statistical mechanics. How do bulk and fine properties arise.  Is there a thermodynamic limit and how is it reached. While the Toda lattice will not show equipartition of energy the FPU system does. The combined system should have a statistical mechanics for any number N particles. 

 Taking a canonical Gibbsian ensemble and partition function is possible, but evaluating the integrals may not be easy. In the region where the Toda is significant , by going to integrals in involution, the exponent in the Gibbs density is replaced by these integrals. However for the FPU part the integrals will have to be evaluated on the energy surfaces that have partial ergodicity ; and the weight function on micropartitions; $ exp(K) $, $K $ is the Kolmogorov entropy.

 The statistical physics of such non exactly solvable systems is not known. However there are a variety of applications of such systems in polymer chains. The dynamical system itself has its interpretation in coupled osmotic cells, corrosive sequence of spots, magnetic and spin lattices etc.  

6. PARTITION FUNCTION FOR HENON HEILES LIKE SYSTEMS AND SPECIFIC HEATS

An easier system to evaluate partly analytically and partly numerically are the Henon Heiles like models, which are taken as model hamiltonians for molecules of a gas or two dimensional domains. A canonical partition function $ \int d \omega  exp (- \beta H )  $ can be integrated by making a partition in constant energy surface . On each such surface the dynamical surface shows coexisting integrable  and chaotic regions.

 The area fraction for these regions as a function of energy is known from computation. It is a straight line function between $ E = 0.11$ and $ E= 0.167 $. The Kolmogorov entropy as a function of energy is known to rise to saturation. This can be put into the additional weight or measure as $ exp (K) $ in the integral. K is zero for integrable or KAM torii regions.  The sum over all regions for each energy surface and the integral over all energies can be evaluated numerically, by splitting $ E==0 $ to $ E= 0.11$ for integrable; then $ E=0.11$ to $ E= 0.167 $ for the mixed regions and $ E= 0.167 $ to $ E = 1 $ for the chaotic case.

The general Hamiltonian of Henon Heiles type is more useful to study the parametric transition to chaos, and hence the dependence of the partition function on these parameters. If an ensemble of these H-H systems is taken , that is a gas of molecules with their internal hamiltonians as H-H, then the chaotic transitions internally can be generated by collisional transfer of energy among molecules. 

If the average energy per particle is less than the critical energy $ E = 0.11$ , then the integrable regions dominate . But as the thermal energy per particle crosses the $ E=0.11$ to $ E= 0.167 $; then all chaotic regions dominate. Hence a $ k T < 0.11$ increasing to $ k T > 0.167 $ change will create a discontinuous change in the partition function and consequently in the total internal energy and specific heat of such a gas. This appears as a different kind of phase transition than the usual first and second order ones in thermodynamics. It may be called an intrinsic phase transition.

ACKNOWLEDGEMENT

In November 1973 I was concerned with the issue of dynamical systems and statistical mechanics, but did not develop my work and paper on ergodic mechanics further ; following up on Y Sinai, KAM, J Ford and others. I thank W. C. Schieve, I. Prigogine and L.Reichl at University of Texas at Austin for early interest in this work. The question of defining partition functions inclusive of K entropy measure remained. 

Many developments occured in classical and quantum chaos over the years, but there was still no statistical mechanics based on them. Then in the years from 1987 to 1996 I attempted to work on simple dynamical systems with implications in statistical mechanics. I thank Physics department, Mumbai university,  School of Physics, Central University , Birla science center and Prof Mondal at Hyderabad, Non Linear dynamics group at NCL, Pune and Raman Research Institute, Bangalore for facilities to work and to present the work in those years. 

It was too early still to successfully publish any work in this field in a journal as no subject classification for it existed till 1998. Then in the 1990s  there was growing work in nano physics, and clusters being published. A renewed interest in the foundations of Statistical mechanics based on dynamical systems and for small or finite N systems  has been seen in 2000 onwards publication. 

Hence it is in acknowledgement of these developments, followed over the years, that I am submitting  short papers on this topic to arXiv. More work is definitely required in this subject to have a complete and final theory.

I thank the Institute of mathematical sciences , Chennai for its facilities; its Director and Dr H Sharatchandra for supporting my visit, and its faculty for discussions.

REFERENCES

1. Ajay Patwardhan arxiv cond-mat 0511231,0411176

2. R Hall P. Wolnyes arxiv 0707.1854

3. V Ilyin , I Procaccia et al arxiv 0705.1043

4. I Grigorenko, S Haas, I Levi arxiv cond-mat 0607252

5. M Pettini, L Cassetti, E.G.D. Cohen et al,arxiv cond -mat

              0410282, 0303200, 0104267, 9912092, 9608054

6. B Gershgorin arxiv 0707.2830

7. G Bermann, F Izrailev, arxiv nlin 0411062

8. V Bulitko arxiv math 0407116

9. L Reichl, A modern course in Statistical Physics , Arnold 1992

10.R Schaeffer H Stoeckmann et al arxiv nlin 0206015

\end{document}